\documentclass[10pt,aps,prd,twocolumn,showpacs,amsmath,nofootinbib]{revtex4-2}

\usepackage[utf8]{inputenc}
\usepackage[american]{babel}
\usepackage[a4paper,margin=2cm]{geometry}

\usepackage{amsmath,amssymb}
\usepackage{booktabs}
\usepackage{graphicx}
\usepackage{caption}
\usepackage{subcaption}

\usepackage{hyperref}

\hypersetup{%
colorlinks,%
citecolor=blue,urlcolor=blue,linkcolor=blue,%
hypertexnames=false,%
pdftitle={Six-pion amplitude},%
pdfauthor={J. Bijnens, T. Husek}%
}

\allowdisplaybreaks

\begin{document}
%%%%%%%%%%%%%%%%%%%%%%%%%%%%%%%%%%%%%%%%%%%%%%%%%%%%%%%%%%%%%%%%%%%%%
% Added for preprint
\onecolumngrid
\thispagestyle{empty}

\begin{flushright}
LU TP 21-30\\[1mm]
Revised August 2021
\end{flushright}

\vskip4cm

\begin{center}
{\huge\bf Six-pion amplitude}\\[3cm]
{\large\bf Johan Bijnens and Tom\'{a}\v{s} Husek}\\[1cm]

{\large Department of Astronomy and Theoretical Physics, Lund University,\\[2mm] S\"olvegatan 14A, SE 223-62 Lund, Sweden}\\[4cm]

{\bf Abstract}
\end{center}

Within the framework of the massive O($N$) nonlinear sigma model extended to the next-to-leading order in the chiral counting (for $N=3$ corresponding to the two(-quark)-flavor Chiral Perturbation Theory), we calculate the relativistic six-pion scattering amplitude at low energy up to and including terms $\mathcal{O}(p^4)$.
Results for the pion mass, decay constant and the four-pion amplitude in the case of $N$ (meson) flavors at $\mathcal{O}(p^4)$ are also presented.

\setcounter{page}{0}
\newpage
\twocolumngrid
%%%%%%%%%%%%%%%%%%%%%%%%%%%%%%%%%%%%%%%%%%%%%%%%%%%%%%%%%%%%%%%%%%%%%

\title{Six-pion amplitude}
\author{Johan Bijnens}
\email{johan.bijnens@thep.lu.se}
\author{Tom\'{a}\v{s} Husek}
\email{tomas.husek@thep.lu.se}
\affiliation{Department of Astronomy and Theoretical Physics, Lund University,\\ S\"olvegatan 14A, SE 223-62 Lund, Sweden}
\date{\today}

\begin{abstract}
Within the framework of the massive O($N$) nonlinear sigma model extended to the next-to-leading order in the chiral counting (for $N=3$ corresponding to the two(-quark)-flavor Chiral Perturbation Theory), we calculate the relativistic six-pion scattering amplitude at low energy up to and including terms $\mathcal{O}(p^4)$.
Results for the pion mass, decay constant and the four-pion amplitude in the case of $N$ (meson) flavors at $\mathcal{O}(p^4)$ are also presented.
\end{abstract}


\maketitle

\section{Introduction}

The theory of the strong interaction is Quantum Chromodynamics (QCD), which due to its underlying non-Abelian structure becomes nonperturbative at low energies. Hence, in this region, it is impractical to study the related effects directly within the fundamental theory and alternative approaches have to be explored. Chiral perturbation theory (ChPT) \cite{Weinberg:1978kz,Gasser:1983yg} is the effective field theory that can be used to study interactions of hadrons in the low-energy domain.
Many observables are known in ChPT to a high loop order, but only recently it has become of interest to calculate the six-pion amplitude at low energies after it was estimated using lattice QCD, as shortly discussed further below.
Therefore, a calculation to next-to-leading order (NLO) in ChPT shall be useful.

The six-pion amplitude at tree level was first done using current algebra methods; see e.g.\ Ref.~\cite{Osborn:1969ku}. It has been redone with Lagrangian methods many times; see Ref.~\cite{Bijnens:2019eze} and references therein. Until now, it has not been known to one-loop order; the tree-level result at NLO was done before in Refs.~\cite{Low:2019ynd,Bijnens:2019eze}.

Two-flavor ChPT is equivalent to a massive O(4)/O(3) nonlinear sigma model extended to higher orders in the expansion.
Furthermore, there has been a recent development on the structure of amplitudes, also for scalar theories; see e.g.\ Refs.~\cite{Cheung:2015ota,Cheung:2016drk,Bijnens:2019eze,Low:2019ynd,Kampf:2019mcd,Low:2020ubn}.
We have therefore calculated at NLO the six-pion amplitude, as well as the four-pion amplitude, pion mass and decay constant for the ChPT generalization to the O($N+1$)/O($N$) massive nonlinear sigma model, which might be useful beyond the QCD two-flavor setting. For three or more flavors one needs to consider the symmetry breaking pattern $\text{SU}(N_\text{f})\times\text{SU}(N_\text{f})/\text{SU}(N_\text{f})$, with $N_\text{f}$ standing for the number of flavors.
The six-pion amplitude in this case is a possible direction for future work.

In this paper, we discuss the calculation and the result of the six-pion amplitude.
Because of the large number of kinematic invariants in a six-pion amplitude, the reduction to master integrals (i.e.\ the scalar triangle integrals) leads to an enormous expression. We have chosen a redundant basis of integrals that have good symmetry properties allowing us to write the amplitude in a fairly compact form.

Three-pion systems corresponding to the six-pion amplitude have been extensively studied in lattice QCD \cite{Mai:2018djl,Blanton:2019vdk,Mai:2019fba,Culver:2019vvu,Fischer:2020jzp,Hansen:2020otl,Brett:2021wyd,Blanton:2021llb}. The relation of our calculation to the measurement on the lattice is nontrivial to implement given the complexity of the three-body finite volume calculations and the subtraction of the two-body rescatterings involved. The main original and some later works are Refs.~\cite{Hansen:2014eka,Hansen:2015zga,Hammer:2017uqm,Hammer:2017kms,Mai:2017bge,Blanton:2020jnm,Romero-Lopez:2020rdq}.
There are also recent reviews introducing the subject and containing many more references \cite{Hansen:2019nir,Mai:2021lwb}.

In Sect.~\ref{sec:theory} we define the O($N+1$)/O($N$) massive nonlinear sigma model as an extension of two-flavor ChPT. We describe the model, its renormalization at NLO and the use of more than one parametrization as a check on our results. The four-pion amplitude, pion mass and decay constant are derived in Sect.~\ref{sec:four-pion}. Our main result, the six-pion amplitude, is discussed in Sect.~\ref{sec:six-pion}, both its calculation and form. Most of the explicit expressions are relegated to App.~\ref{app:6pi:res}. Some examples of numerical results are given in Sect.~\ref{sec:numerical} and our conclusions are shortly discussed in Sect.~\ref{sec:conclusions}. Finally, App.~\ref{app:integrals} discusses the integrals we use.

The analytical work in this manuscript was done both using Wolfram {\em Mathematica} with \textsc{FeynCalc} package \cite{Mertig:1990an,Shtabovenko:2016sxi,Shtabovenko:2020gxv} and a \textsc{FORM}~\cite{Vermaseren:2000nd} implementation.
The numerical results use \textsc{LoopTools}~\cite{vanOldenborgh:1989wn,Hahn:1998yk}.

\section{Theoretical setting}
\label{sec:theory}

The massive O($N+1$)/O($N$) nonlinear sigma model Lagrangian reads
\begin{equation}
\mathcal{L}_{\text{nL}\sigma\text{M}}
=\frac{F^2}2\,\partial_\mu\Phi^\mathsf{T}\partial^\mu\Phi+F^2\chi^\mathsf{T}\Phi\,,
\label{eq:Lns}
\end{equation}
with $\Phi$ being a real vector of $N+1$ components, which transforms as the fundamental representation of O($N+1$) and satisfies $\Phi^\mathsf{T}\Phi=1$.
The spontaneous symmetry breaking of O($N+1$) to O($N$) is triggered by the vacuum, which takes the form
\begin{equation}
\big\langle\Phi^\mathsf{T}\big\rangle
=\left(1,\,\vec 0\,\right).
\end{equation}
The term in Eq.~\eqref{eq:Lns} containing $\chi$ explicitly breaks the symmetry to the same O($N$), setting
\begin{equation}
\chi^\mathsf{T}
=\left(M^2,\,\vec 0\,\right).
\end{equation}
Above, $F$ and $M$ are the bare pion decay constant and mass, respectively.
For $N=3$ meson flavors, the Lagrangian \eqref{eq:Lns} corresponds to the lowest-order Lagrangian of two(-quark)-flavor Chiral Perturbation Theory.
This can be extended beyond the leading order (LO) in the following way:
\begin{equation}
\begin{split}
\mathcal{L}
&=\mathcal{L}_{\text{nL}\sigma\text{M}}\\
&+l_1\big(\partial_\mu\Phi^\mathsf{T}\partial^\mu\Phi\big)\big(\partial_\nu\Phi^\mathsf{T}\partial^\nu\Phi\big)\\
&+l_2\big(\partial_\mu\Phi^\mathsf{T}\partial_\nu\Phi\big)\big(\partial^\mu\Phi^\mathsf{T}\partial^\nu\Phi\big)\\
&+l_3\big(\chi^\mathsf{T}\Phi\big)^2
+l_4\partial_\mu\chi^\mathsf{T}\partial^\mu\Phi\,.
\end{split}
\label{eq:L}
\end{equation}
We use the above Lagrangian to calculate the amplitudes in question at the NLO.
External fields, as needed for the decay constant, can be added as in Ref.~\cite{Gasser:1983yg}.
The coefficients (low-energy constants) $l_i$ are free parameters in the theory, and carry both UV divergent and finite parts expressed as
\begin{equation}
l_i
=(c\mu)^{d-4}\left(\frac1{16\pi^2}\frac1{d-4}\,\gamma_i+l_i^\text{r}\right).
\label{eq:li}
\end{equation}
Above, $c$ is such that
\begin{equation}
\log c
=-\frac12\left(1-\gamma_\text{E}+\log4\pi\right).
\end{equation}
Consequently, in terms of $\epsilon=2-d/2$ and
\begin{equation}
\frac1{\tilde\epsilon}
\equiv\frac{1}{\epsilon}-\gamma_\text{E}+\log4\pi-\log\mu^2+1\,,
\label{eq:epstilde}
\end{equation}
one writes
\begin{equation}
l_i
=-\kappa\,\frac{\gamma_i}2\frac1{\tilde\epsilon}+l_i^\text{r}\,,
\label{eq:li2}
\end{equation}
with $\kappa\equiv1/(16\pi^2)$.
Notice the extra `+1' term in Eq.~\eqref{eq:epstilde} with respect to the $\overline{\text{MS}}$ renormalization scheme.
When considering the four-pion amplitude at the order $\mathcal{O}(p^4)$, the coefficients $\gamma_i$ are uniquely fixed by the requirement of the cancellation of the divergent parts among the NLO tree-level (containing vertices with $l_i$s) and one-loop contributions.
From studying the pion mass, decay constant and the four-pion amplitude discussed in the following section we find
\begin{equation}
\begin{split}
\gamma_1&=\frac N2-\frac76\,,\\
\gamma_2&=\frac23\,,\\
\gamma_3&=1-\frac N2\,,\\
\gamma_4&=N-1\,.
\end{split}
\end{equation}

To expand the Lagrangian~\eqref{eq:L} in terms of the pion fields $\phi_i$, $i=1,\dots,N$, one needs to use a particular representation for $\Phi$.
Using multiple representations simultaneously then serves --- together with the cancellation of the UV divergent parts --- as the fundamental cross-check pointing to the validity of the final expression, since, of course, one expects to obtain the same (physical) result irrespective of the parametrization used in the intermediate steps.
We used the following five representations:
\begin{align}
\Phi_1
&=\left(\sqrt{1-\varphi},\,\frac{\pmb{\phi}^\mathsf{T}}F\right)^\mathsf{T}\,,\\
\Phi_2
&=\frac1{\sqrt{1+\varphi}}\left(1,\,\frac{\pmb{\phi}^\mathsf{T}}F\right)^\mathsf{T}\,,\\
\Phi_3
&=\left(1-\frac12\varphi,\,\sqrt{1-\frac14\varphi}\,\frac{\pmb{\phi}^\mathsf{T}}F\right)^\mathsf{T}\,,\\
\Phi_4
&=\left(\cos\sqrt{\varphi},\,\frac1{\sqrt{\varphi}}\sin\sqrt{\varphi}\,\frac{\pmb{\phi}^\mathsf{T}}F\right)^\mathsf{T}\,,\\
\Phi_5
&=\frac1{1+\frac14\varphi}\left(1-\frac14\varphi,\,\frac{\pmb{\phi}^\mathsf{T}}F\right)^\mathsf{T}\,.
\end{align}
Above, we denote $\varphi\equiv\frac{\pmb{\phi}^\mathsf{T}\pmb{\phi}}{F^2}$, with $\pmb{\phi}^\mathsf{T}=(\phi_1,\dots,\phi_N)$ being a real vector of $N$ components (flavors) transforming linearly under the unbroken part of the O($N$) symmetry group.
The parametrization $\Phi_1$ is the one used by Gasser and Leutwyler in Ref.~\cite{Gasser:1983yg}, $\Phi_2$ is a simple variation, $\Phi_3$ is such that the term explicitly breaking the symmetry in Eq.~\eqref{eq:Lns} only gives mass terms of $\phi_i$s but no vertices, $\Phi_4$ represents the result when one follows the general prescription from Ref.~\cite{Coleman:1969sm} and $\Phi_5$ is the one originally introduced by Weinberg~\cite{Weinberg:1968de}.
Finally, these are just a few examples of the whole class of parametrizations that keep the O($N+1$) symmetry manifest:
\begin{equation}
\Phi
=\left(\sqrt{1-\varphi\,f^2(\varphi)},\,f(\varphi)\,\frac{\pmb{\phi}^\mathsf{T}}F\right)^\mathsf{T}\,.
\end{equation}
Above, $f(x)$ is any analytical function satisfying $f(0)=1$.

\section{Four-pion amplitude}
\label{sec:four-pion}

Let us start with the four-pion amplitude $A_{4\pi}$, assuming $N$ flavors of pseudoscalar mesons (pions), since this result has (to our knowledge) not been yet presented in the literature.
We write the four-pion amplitude exactly in the form as given in Refs.~\cite{Bijnens:1995yn,Bijnens:1997vq}, generalized to $N\ne3$.
Note that this is somewhat different from the form given in Ref.~\cite{Gasser:1983yg}:
Both results are, of course, equivalent to the given order $\mathcal{O}(p^4)$, but lead to different off-shell extrapolations.

In general, for the {\em on-shell} amplitude with all the pion incoming 4-momenta $p_i$ and flavors $f_i$, $i=1,\dots,4$, $\sum_i p_i=0$, we can write due to invariance under rotation in the isospin space and crossing symmetry
\begin{equation}
\begin{split}
&A_{4\pi}(p_1,f_1,p_2,f_2,p_3,f_3,f_4)\\
  &=\delta_{f_1f_2}\delta_{f_3f_4}A(p_1,p_2,p_3)\\
  &+\delta_{f_1f_3}\delta_{f_2f_4}A(p_3,p_1,p_2)\\
  &+\delta_{f_2f_3}\delta_{f_1f_4}A(p_2,p_3,p_1)\,.
\end{split}
\label{eq:A4pi}
\end{equation}
In terms of the Mandelstam variables $s=(p_1+p_2)^2$, $t=(p_1+p_3)^2$ and $u=(p_2+p_3)^2$, $s+t+u=4M^2$, one has for the subamplitude $A(p_1,p_2,p_3)=A(s,t,u)$.
The latter can be written up to and including $\mathcal{O}(p^4)$, collecting the contributions order by order, as
\begin{equation}
A(s,t,u)
=A^{(2)}(s,t,u)+A^{(4)}(s,t,u)\,.
\end{equation}

\begin{figure}[t]
\includegraphics[width=0.25\columnwidth]{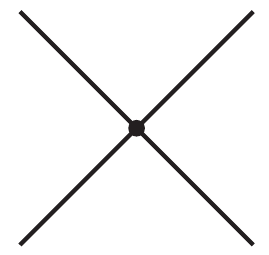}
\caption{
Leading-order contribution to the four-pion amplitude, with the vertex stemming from $\mathcal{L}_{\text{nL}\sigma\text{M}}$.
}
\label{fig:4pi_LO}
\end{figure}

\begin{figure}[!t]
\centering
\begin{subfigure}[t]{0.33\columnwidth}
\includegraphics[width=0.5\columnwidth,angle=90]{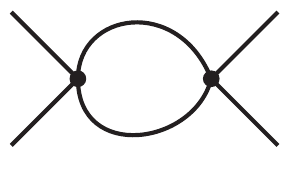}
\caption{3$\times$}
\label{fig:4pi_NLO_bubble}
\end{subfigure}
\begin{subfigure}[t]{0.32\columnwidth}
\includegraphics[width=0.9\columnwidth]{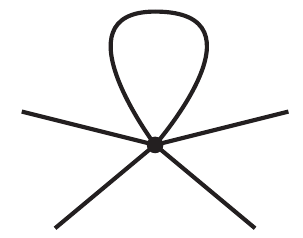}
\caption{1$\times$}
\label{fig:4pi_NLO_tadpole}
\end{subfigure}
\begin{subfigure}[t]{0.32\columnwidth}
\includegraphics[width=0.65\columnwidth]{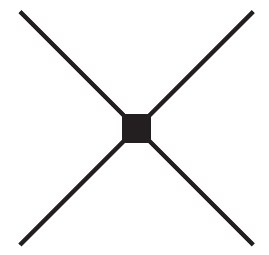}
\caption{1$\times$}
\label{fig:4pi_NLO_CT}
\end{subfigure}
\caption{
NLO topologies relevant for the calculation of the four-pion amplitude.
The multiplicities of the respective diagrams are quoted as subcaptions.
The standard dotlike vertices come from $\mathcal{L}_{\text{nL}\sigma\text{M}}$, while the square vertex stems from the terms proportional to $l_i$s in Eq.~\eqref{eq:L}.
As indicated in the main text, there are other diagrams not explicitly depicted here, relevant at NLO for the field renormalization, and mass and decay-constant redefinitions.
}
\label{fig:4pi_NLO}
\end{figure}

The leading-order tree-level $\mathcal{O}(p^2)$ amplitude stems from a single diagram shown in Fig.~\ref{fig:4pi_LO} (schematically $A_{4\pi}^{(2)}={\mathcal{M}}_\text{LO}^{(2)}|_\text{on-shell}$) and the related subamplitude (with LO relations $M\to M_\pi$ and $F\to F_\pi$) reads
\begin{equation}
A^{(2)}(s,t,u)
=\frac{1}{F_\pi^2}\left(s-M_\pi^2\right).
\label{eq:A4pisub_LO}
\end{equation}
At the next-to-leading order, one has one-loop diagrams (two topologies of four one-loop diagrams in total) combined with a counterterm, as shown in Fig.~\ref{fig:4pi_NLO}, together with NLO field renormalization, and mass and decay-constant redefinitions (at the given order) applied to the LO graph.
Schematically, this can be written as
\begin{equation}
A_{4\pi}^{(4)}
=\mathcal{M}_\text{1-loop}
+\mathcal{M}_\text{CT}
+4(Z^{1/2}-1)\mathcal{M}_\text{LO}^{(2)}
+{\mathcal{M}}_\text{LO}^{(4)}
\,.
\label{eq:A4pi_NLO}
\end{equation}
Note that while the above combination is parametrization independent, the separate terms are not.
The matrix element $\mathcal{M}_\text{1-loop}$ is obtained from the diagrams in Figs.~\ref{fig:4pi_NLO_bubble}, \ref{fig:4pi_NLO_tadpole}, and $\mathcal{M}_\text{CT}$ relates to Fig.~\ref{fig:4pi_NLO_CT}.
The $Z$ factor used for the field renormalization is related to the pion self-energy $\Sigma$ as
\begin{equation}
\frac1Z
=1-\frac{\partial\Sigma(p^2)}{\partial p^2}\bigg|_{p^2=M_\pi^2}\,,
\end{equation}
with $-i\Sigma$ being represented by a tadpole graph with two external legs plus a counterterm stemming from the $l_3$ term in the Lagrangian \eqref{eq:L}.
The matrix element ${\mathcal{M}}_\text{LO}^{(4)}$ denotes the NLO part of the extension of the LO vertex obtained in terms of the replacements
\begin{equation}
\begin{split}
M^2&\to M_\pi^2+\overline\Sigma\,,\\
\frac1{F^2}&\to\frac1{F_\pi^2}(1+2\delta F)\,,
\end{split}
\end{equation}
(at the given order equivalent to standard $M_\pi^2=M^2-\overline\Sigma$, $F_\pi=F(1+\delta F)$), with
\begin{equation}
\begin{split}
\overline\Sigma&=\frac{M_\pi^4}{F_\pi^2}\left[2l_3^\text{r}+\frac12(N-2)L\right]+\mathcal{O}\bigg(\frac1{F_\pi^4}\bigg)\,,\\
\delta F&=\frac{M_\pi^2}{F_\pi^2}\left[l_4^\text{r}-\frac12(N-1)L\right]+\mathcal{O}\bigg(\frac1{F_\pi^4}\bigg)\,.
\end{split}
\end{equation}
Needless to say, in the final result one only retains the terms relevant at the order $\mathcal{O}(p^4)$, i.e.\ terms $\mathcal{O}(1/F_\pi^6)$ are systematically neglected.
Thus, in the rest of the expression \eqref{eq:A4pi_NLO} one simply takes $M\to M_\pi$ and $F\to F_\pi$.
Finally, the NLO part of the subamplitude can be written fairly compactly as of the momenta in the following way:
\begin{align}
&F_\pi^4A^{(4)}(s,t,u)
=(t-u)^2\left(
    -\frac5{36}\,\kappa
    -\frac16\,L
    +\frac12\,l_2^\text{r}
\right)\notag\\
&+M_\pi^2s\bigg[
    \bigg(N-\frac{29}9\bigg)\kappa
    +\bigg(N-\frac{11}3\bigg)L
    -8l_1^\text{r}
    +2l_4^\text{r}
\bigg]\notag\\
&+s^2\bigg[
    \bigg(\frac{11}{12}-\frac N2\bigg)\kappa
    +\bigg(1-\frac N2\bigg)L
    +2l_1^\text{r}
    +\frac12\,l_2^\text{r}
\bigg]\notag\\
&+M_\pi^4 \bigg[
    \bigg(\frac{20}{9}-\frac N2\bigg)\kappa
    +\bigg(\frac83-\frac N2\bigg)L\notag\\
    &\qquad+8\,l_1^\text{r}
    +2l_3^\text{r}
    -2l_4^\text{r}
\bigg]\notag\\
&+\bar J(s)\bigg[
    \bigg(\frac N2-1\bigg) s^2
    + (3-N) M_\pi^2 s
    + \bigg(\frac N2-2\bigg) M_\pi^4
\bigg]\notag\\
&+\bigg\{\frac16\,\bar J(t)\big[
    2 t^2
    - 10 M_\pi^2 t
    - 4 M_\pi^2 s
    + s t
    + 14 M_\pi^4
\big]\notag\\
&\qquad+(t\leftrightarrow u)\bigg\}\,.
\label{eq:A4pisub_NLO}
\end{align}
Above we have used
\begin{equation}
L\equiv\kappa\log\frac{M_\pi^2}{\mu^2}\,.
\end{equation}
The expressions presented in this section agree with the known results for $N=3$~\cite{Gasser:1983yg}, as well as with those on the $N$ dependence (see e.g.\ Refs.~\cite{Dobado:1994fd,Bijnens:2009zi,Bijnens:2010xg}).

\section{Six-pion amplitude}
\label{sec:six-pion}

Before we get to discussing the amplitude itself, let us first talk a bit about combinatorics, since in the case of the six-pion amplitude things get noticeably more complicated compared to the four-pion case where only three different channels/permutations could appear (when distributing four pions in two pairs).
There are ten ways one can distribute the six pions in two groups of three.
We denote these permutations as $P_{10}$ with
\begin{equation}
\begin{split}
  &(i,j,k)(l,m,n)
  =\big\{(1,2,3)(4,5,6),\\
  &\quad(1,2,4)(3,5,6),(1,2,5)(3,4,6),(1,2,6)(3,4,5),\\
  &\quad(1,3,4)(2,5,6),(1,3,5)(2,4,6),(1,3,6)(2,4,5),\\
  &\quad(1,4,5)(2,3,6),(1,4,6)(2,3,5),(1,5,6)(2,3,4)
  \big\}\,.
\end{split}
\end{equation}
Similarly, there are 15 ways the six pions can be distributed in three pairs.
These we denote as $P_{15}$ and are
\begin{equation}
\begin{split}
  &(i,j)(k,l)(m,n)\\
  &=\big\{
  (1,2)(3,4)(5,6),(1,2)(3,5)(4,6),(1,2)(3,6)(4,5),\\
  &(1,3)(2,4)(5,6),(1,3)(2,5)(4,6),(1,3)(2,6)(4,5),\\
  &(1,4)(2,3)(5,6),(1,4)(2,5)(3,6),(1,4)(2,6)(3,5),\\
  &(1,5)(2,3)(4,6),(1,5)(2,4)(3,6),(1,5)(2,6)(3,4),\\
  &(1,6)(2,3)(4,5),(1,6)(2,4)(3,5),(1,6)(2,5)(3,4)
  \big\}\,.
\end{split}
\end{equation}

We define a general six-pion amplitude with all the pion incoming 4-momenta $p_i$ and flavors $f_i$, $i=1,\dots,6$, so it is a function of $p_1,f_1$,\,\dots,\,$p_6,f_6$.
The full six-pion amplitude at $\mathcal{O}(p^4)$ can be written as
\begin{equation}
A_{6\pi}
=A_{6\pi}^{(4\pi)}+A_{6\pi}^{(6\pi)}\,.
\label{eq:A6piparts}
\end{equation}
Above, $A_{6\pi}^{(4\pi)}$ is the part that can be written in terms of the four-pion amplitude and $A_{6\pi}^{(6\pi)}$ is the remainder.
The first part contains a single pole and is of the form
\begin{equation}
\begin{split}
A_{6\pi}^{(4\pi)}
&\equiv\sum_{P_{10},\,f_\text{o}}
A_{4\pi}(p_i,f_i,p_j,f_j,p_k,f_k,f_\text{o})\\
&\times\frac{(-1)}{p_{ijk}^2-M^2_\pi}\,
A_{4\pi}(p_l,f_l,p_m,f_m,p_n,f_n,f_\text{o})\,,
\end{split}
\label{eq:A6pi4pi}
\end{equation}
with $f_\text{o}$ being the flavor of the internal propagator and $p_{ijk}\equiv p_i+p_j+p_k$.
Above, $A_{4\pi}(p_i,f_i,p_j,f_j,p_k,f_k,f_\text{o})$ is the four-pion amplitude with one leg off-shell.
Similarly to Eq.~\eqref{eq:A4pi}, we write
\begin{equation}
\begin{split}
  &A_{4\pi}(p_i,f_i,p_j,f_j,p_k,f_k,f_\text{o})\\
  &=\delta_{f_if_j}\delta_{f_kf_\text{o}}A(p_i,p_j,p_k)\\
  &+\delta_{f_if_k}\delta_{f_jf_\text{o}}A(p_k,p_i,p_j)\\
  &+\delta_{f_jf_k}\delta_{f_if_\text{o}}A(p_j,p_k,p_i)\,.
\end{split}
\label{eq:A4pi4pi}
\end{equation}
The (four-pion) subamplitude $A(p_i,p_j,p_k)=A(s,t,u)$ is defined as usual, with $s=(p_i+p_j)^2$, $t=(p_i+p_k)^2$ and $u=(p_j+p_k)^2$.
However, note that these variables now satisfy $s+t+u=3M^2_\pi+p_{ijk}^2$.
In Eq.~\eqref{eq:A6pi4pi}, the residue at the pole is unique (related to the uniqueness of the {\em on-shell} four-pion amplitude), however, the off-shell extrapolation away from $p_{ijk}^2=M^2_\pi$ is not and is subject to a choice determining how the particular contributions are redistributed among the two parts of Eq.~\eqref{eq:A6piparts}.
This split into the factorizable ($A_{6\pi}^{(4\pi)}$) and nonfactorizable ($A_{6\pi}^{(6\pi)}$) parts is convenient since there are momentum configurations where the intermediate propagators (appearing subsequently only in $A_{6\pi}^{(4\pi)}$) can become on-shell.
At the same time, as discussed later on, this structure naturally shows up diagrammatically at NLO.
It is then anticipated that --- going beyond the six-particle amplitude --- similar comments apply: Residues at poles are unique but away from the poles there are many more ambiguities.

For the subamplitude $A(s,t,u)$ we choose the form as given by Eqs.~\eqref{eq:A4pisub_LO} and \eqref{eq:A4pisub_NLO}.
Other off-shell extrapolations are possible and will lead to a different $A_{6\pi}^{(6\pi)}$, which has no poles, only cuts,%
\footnote{This is not quite true: The imaginary part of the triangle integrals can contain poles.}
and can be written in the form
\begin{equation}
\label{eq:A6piP15}
A_{6\pi}^{(6\pi)}
\equiv\sum_{P_{15}}\delta_{f_if_j}\delta_{f_kf_l}\delta_{f_mf_n}A(p_i,p_j,p_k,p_l,p_m,p_n)\,.
\end{equation}
The (six-pion) subamplitude $A(p_1,p_2,p_3,p_4,p_5,p_6)$ should be thought of
as a function of three pairs of momenta. It is fully symmetric under the interchange
of any of the pairs as well as symmetric for the interchange within a pair.
We have chosen a particular form for the {\em off-shell} four-pion subamplitude $A(s,t,u)$, independent of the parametrization used, so the amplitude $A_{4\pi}$ and the respective parts $A_{6\pi}^{(4\pi)}$ and $A_{6\pi}^{(6\pi)}$ from Eq.~\eqref{eq:A6piparts} are, as a consequence, parametrization independent.
However, the way the contributions from the one-particle irreducible and reducible diagrams are distributed within the final result is parametrization dependent.

\begin{figure}[t]
\centering
\begin{subfigure}[t]{0.45\columnwidth}
\includegraphics[width=0.8\columnwidth]{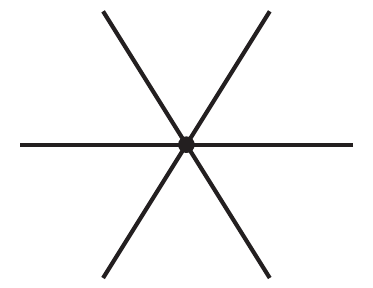}
\caption{1$\times$}
\label{fig:6pi_LO_a}
\end{subfigure}
\begin{subfigure}[t]{0.5\columnwidth}
\includegraphics[width=0.9\columnwidth]{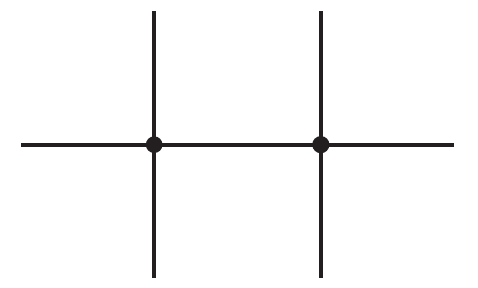}
\caption{10$\times$}
\label{fig:6pi_LO_b}
\end{subfigure}
\caption{
Leading-order contributions to the six-pion amplitude.
The topology depicted on the right in the subfigure (b) manifests itself with ten contributions with different permutations of the external legs.
}
\label{fig:6pi_LO}
\end{figure}

\begin{figure}[!t]
\centering
\begin{subfigure}[t]{0.33\columnwidth}
\includegraphics[width=0.9\columnwidth]{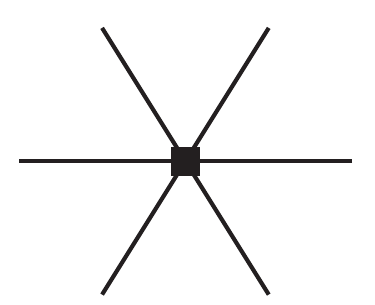}
\caption{1$\times$}
\label{fig:6piNLOa}
\end{subfigure}
\begin{subfigure}[t]{0.32\columnwidth}
\includegraphics[width=0.9\columnwidth]{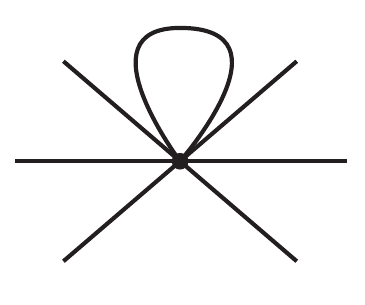}
\caption{1$\times$}
\label{fig:6piNLOb}
\end{subfigure}
\begin{subfigure}[t]{0.32\columnwidth}
\includegraphics[width=0.9\columnwidth]{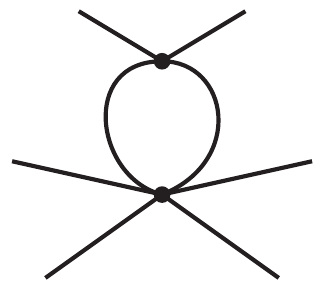}
\caption{15$\times$}
\label{fig:6piNLOc}
\end{subfigure}

\vspace{2mm}
\begin{subfigure}[t]{0.33\columnwidth}
\includegraphics[width=0.9\columnwidth]{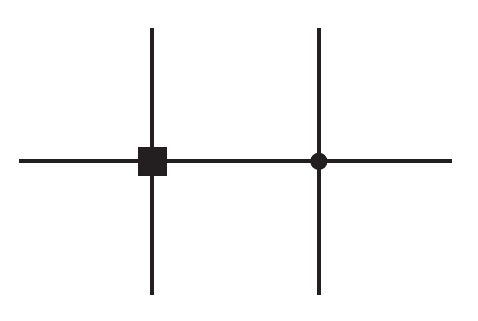}
\caption{20$\times$}
\label{fig:6piNLOd}
\end{subfigure}
\begin{subfigure}[t]{0.32\columnwidth}
\includegraphics[width=0.9\columnwidth]{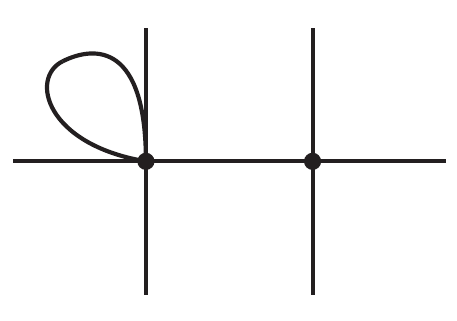}
\caption{20$\times$}
\label{fig:6piNLOe}
\end{subfigure}
\begin{subfigure}[t]{0.32\columnwidth}
\includegraphics[width=0.9\columnwidth]{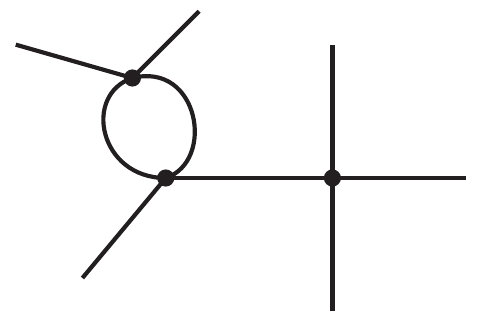}
\caption{60$\times$}
\label{fig:6piNLOf}
\end{subfigure}

\vspace{2mm}
\begin{subfigure}[t]{0.33\columnwidth}
\includegraphics[width=0.9\columnwidth]{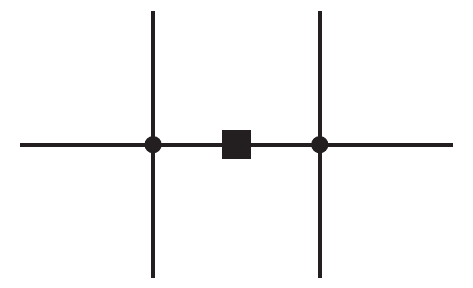}
\caption{10$\times$}
\label{fig:6piNLOg}
\end{subfigure}
\begin{subfigure}[t]{0.32\columnwidth}
\includegraphics[width=0.9\columnwidth]{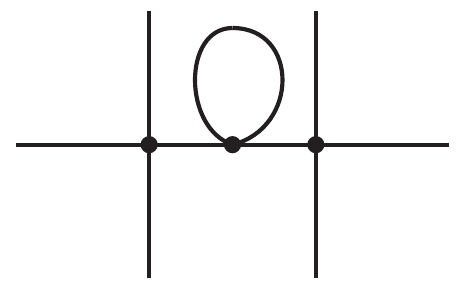}
\caption{10$\times$}
\label{fig:6piNLOh}
\end{subfigure}
\begin{subfigure}[t]{0.32\columnwidth}
\includegraphics[width=0.9\columnwidth]{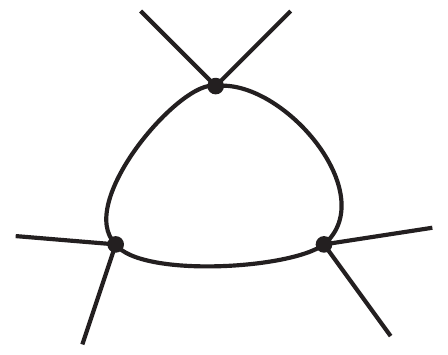}
\caption{15$\times$}
\label{fig:6piNLOi}
\end{subfigure}
\caption{
NLO topologies relevant for the six-pion amplitude.
The multiplicities of the respective diagrams are quoted as subcaptions.
}
\label{fig:6piNLO}
\end{figure}

Regarding the six-pion amplitude at the leading order, there are 1 + 10 tree diagrams depicted in Fig.~\ref{fig:6pi_LO}.
The diagram in Fig.~\ref{fig:6pi_LO_a} only contributes to $A_{6\pi}^{(6\pi)}$, but the one in Fig.~\ref{fig:6pi_LO_b} contributes to both the pole and nonpole parts $A_{6\pi}^{(4\pi)}$ and $A_{6\pi}^{(6\pi)}$, respectively.
At the next-to-leading order, the one-particle irreducible (1PI) diagrams shown in Figs.~\ref{fig:6piNLOa}, \ref{fig:6piNLOb}, \ref{fig:6piNLOc} and \ref{fig:6piNLOi} clearly contribute only to $A_{6\pi}^{(6\pi)}$, together with the field, mass and decay-constant renormalizations at NLO applied to the LO expression stemming from the first LO graph (in Fig.~\ref{fig:6pi_LO_a}).
Schematically,
\begin{equation}
\begin{split}
&A_{6\pi}^{(6\pi)}\big|_\text{NLO,1PI}\\
&=\tilde{\mathcal{M}}_\text{1-loop}^{(6\pi)}
+\tilde{\mathcal{M}}_\text{CT}^{(6\pi)}
+6(Z^{1/2}-1)\tilde{\mathcal{M}}_\text{LO}^{(a)(2)}
+\tilde{\mathcal{M}}_\text{LO}^{(a)(4)}
\,.
\end{split}
\label{eq:6pi_A4}
\end{equation}
The discussion of the pole part is somewhat more complicated.
The double-pole part of the contributions represented by diagrams depicted in Figs.~\ref{fig:6piNLOg} and \ref{fig:6piNLOh} cancels with the piece due to the propagator mass renormalization in the LO pole contribution, and consequently
\begin{equation}
\begin{split}
&\tilde{\mathcal{M}}^\text{2-pole}
+6(Z^{1/2}-1)\tilde{\mathcal{M}}_\text{LO}^{(b)(2)}
+\tilde{\mathcal{M}}_\text{LO}^{(b)(4)}\big|_\text{prop.}\\
&=8(Z^{1/2}-1)\tilde{\mathcal{M}}_\text{LO}^{(b)(2)}
\,.
\end{split}
\end{equation}
This, together with $\tilde{\mathcal{M}}_\text{LO}^{(b)(2)}+\tilde{\mathcal{M}}_\text{LO}^{(b)(4)}\big|_\text{vert.}$ and contributions stemming from the topologies shown in Figs.~\ref{fig:6piNLOd}, \ref{fig:6piNLOe} and \ref{fig:6piNLOf} gives, schematically, the equivalent of two NLO $\pi\pi$ scatterings connected with the propagator, i.e.\ precisely the structure of Eq.~\eqref{eq:A6pi4pi}.
Choosing the particular off-shell form of $A_{6\pi}^{(4\pi)}$ as discussed earlier, the remainder is deferred to $A_{6\pi}^{(6\pi)}$.

Finally, let us present the results.
The six-pion subamplitude can be again written in the following form respecting orders in the expansion we use:
\begin{equation}
\label{eq:defA}
\begin{split}
A(p_1,p_2,p_3,p_4,p_5,p_6)
&=A^{(2)}(p_1,p_2,p_3,p_4,p_5,p_6)\\
&+A^{(4)}(p_1,p_2,p_3,p_4,p_5,p_6)\,.
\end{split}
\end{equation}
At the leading order we find a simple expression
\begin{equation}
\label{eq:defA2}
\begin{split}
&A^{(2)}(p_1,p_2,p_3,p_4,p_5,p_6)\\
&=\frac{1}{F_\pi^4}\left(2p_1\cdot p_2+2p_3\cdot p_4+2p_5\cdot p_6+3M_\pi^2\right)\\
&=\frac{1}{F_\pi^4}\left(q_1^2+q_2^2+q_3^2-3M_\pi^2\right).
\end{split}
\end{equation}
Above we used $q_1=p_1+p_2$, $q_2=p_3+p_4$, $q_1=p_5+p_6$.
Note that the dependence on momenta is the only one at this order compatible with the symmetries of the amplitude.
This expression agrees with known results; see e.g.\ Refs.~\cite{Osborn:1969ku,Bijnens:2019eze,Blanton:2019vdk}.

The main new result presented in this work is the next-order six-pion subamplitude.
We split it up into numerous parts:
\begin{equation}
\label{eq:defA4}
\begin{split}
&F_\pi^6 A^{(4)}(p_1,p_2,\dots,p_6)
=A_{C_3}+A_{C_{21}}^{(1)}+A_{C_{21}}^{(2)}+A_{C_{11}}\\
&+A_C^{(1)}+A_C^{(2)}+A_C^{(3)}+A_J^{(1)}+A_J^{(2)}+A_\pi+A_L+A_l\,.
\end{split}
\end{equation}
Each of the terms on the right-hand side has the required symmetries under interchange of momenta.
In the above expression, we suppressed the arguments $(p_1,p_2,\dots,p_6)\equiv(p_1,p_2,p_3,p_4,p_5,p_6)$.
Not to break the flow of the paper, we moved the results for each of the above parts to Appendix~\ref{app:6pi:res}.

\section{Numerical results}
\label{sec:numerical}

We only present a few numerical results here since the full analysis of the finite volume and the subtraction of the two-body rescatterings is very nontrivial; see Refs.~\cite{Hansen:2019nir,Mai:2021lwb} and references therein.

We choose a symmetric three to three scattering configuration given by
\begin{align}
\label{eq:kinematics}
    p_1&=\left(E_p,p,0,0\right),\notag\\
    p_2&=\left(E_p,-\frac{1}{2}p,\frac{\sqrt{3}}{2}p,0\right),\notag\\
    p_3&=\left(E_p,-\frac{1}{2}p,-\frac{\sqrt{3}}{2}p,0\right),\notag\\
    p_4&=\left(-E_p,0,0,p\right),\notag\\
    p_5&=\left(-E_p,\frac{\sqrt{3}}{2}p,0,-\frac{1}{2}p\right),\notag\\
    p_6&=\left(-E_p,-\frac{\sqrt{3}}{2}p,0,-\frac{1}{2}p\right),
\end{align}
with $E_p=\sqrt{p^2+M^2}$.
The numerical inputs we use are
\begin{align}
    M_\pi&=0.139570\,\text{GeV}\,,&  \bar l_1&=-0.4\,,\notag\\
    F_\pi&=0.0927\,\text{GeV}\,,&  \bar l_2&=4.3\,, \notag\\
    \mu&=0.77\,\text{GeV}\,,&  \bar l_3&=3.41\,,\notag\\
    N&=3\,,&  \bar l_4&=4.51\,,
\end{align}
where the values for $\bar l_i$ are from Refs.~\cite{Bijnens:2014lea,Colangelo:2001df,Aoki:2016frl}.
One then obtains $l_i^\text{r}$ (introduced in Eqs.~\eqref{eq:li}, \eqref{eq:li2}) appearing in our results after employing $l_i^r = \frac12\kappa\gamma_i\big(\bar l_i + \ln(M_\pi^2/\mu^2)\big)$, as defined in Ref.~\cite{Gasser:1983yg}.

\begin{figure}[tb]
    \centering
    \includegraphics[width=0.99\columnwidth]{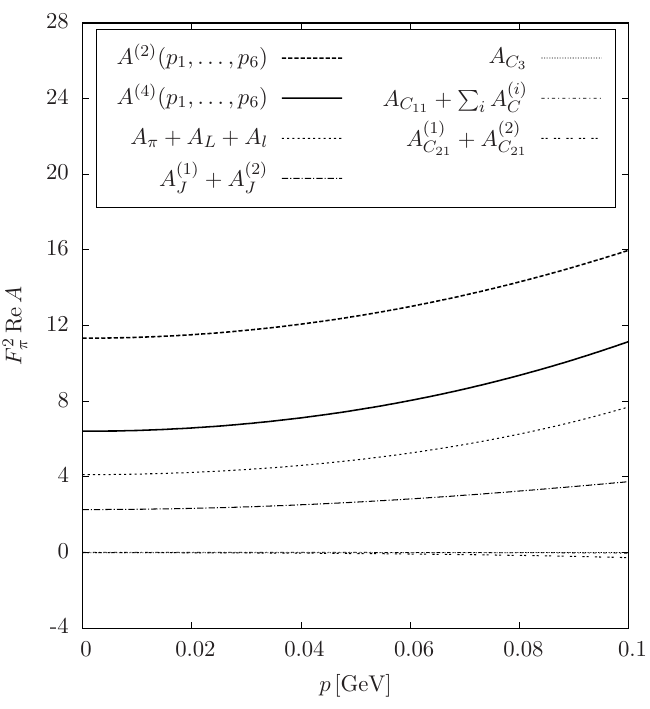}
\caption{
    The six-pion subamplitude in the kinematic configuration of Eq.~\eqref{eq:kinematics}.
    Plotted are the lowest-order result $A^{(2)}(p_1,p_2,\dots,p_6)$ of Eq.~\eqref{eq:defA2} and the next-to-leading-order result $A^{(4)}(p_1,p_2,\dots,p_6)$ of Eq.~\eqref{eq:defA4} together with several groups of its individual constituents.
    We show only the real parts.
    To enhance readability, the order in the key is consistent with how the curves appear in the plot.
    Following Eq.~\eqref{eq:defA4}, the solid curve representing the NLO result is the sum of all the curves below it.
    The lines corresponding to $A_{C_3}$ and $A_{C_{11}}+\sum_i A_C^{(i)}$ are added for completeness; in this plot, they lie very close to zero and basically overlap.
}
    \label{fig:plotA}
\end{figure}

The six-pion subamplitude as introduced in Eq.~\eqref{eq:defA} is plotted in Fig.~\ref{fig:plotA}.
As one can see, the corrections are sizable but not enormous.
They are, however, much smaller than the part coming from the four-pion amplitude as given in Eq.~\eqref{eq:A6pi4pi}.
To compare the same flavor structures, we combine Eqs.~\eqref{eq:A6pi4pi} and \eqref{eq:A4pi4pi} and rewrite the result in the form of Eq.~\eqref{eq:A6piP15}.
The size of the different contributions for $p=0.1\,\text{GeV}$ (the end point of the shown curves) is given in Tab.~\ref{tab:numerics}.
\begin{table}[tb]
    \centering
    \begin{tabular}{c|r||c|r}
    \toprule
    \multicolumn{4}{c}{$F_\pi^2\times\operatorname{Re}A$} \\
    \midrule
        $A^{(4\pi)}_{6\pi}$ (LO) & \,$-319.00$ & \,$A^{(2)}(p_1,\dots,p_6)$\, & 15.99 \\
        $A^{(4\pi)}_{6\pi}$ (NLO)\, & $-28.54$ & \,$A^{(4)}(p_1,\dots,p_6)$\, & 11.16 \\
    \midrule\midrule
    \multicolumn{4}{c}{$F_\pi^2\times\operatorname{Re}A/F_\pi^6$} \\
    \midrule
        $A_{C_3}$  & 0.002 & $A_{J}^{(1)}$  & 1.917 \\
        $A_{C_{21}}^{(1)}$  & $-0.948$ & $A_{J}^{(2)}$  & 1.835 \\
        $A_{C_{21}}^{(2)}$  & 0.682 & $A_\pi$ & \,$-2.488$ \\
        $A_{C_{11}}$  & 0.090 & $A_L$ & 8.985 \\
        $A_{C}^{(1)}$  & $-0.026$ & $A_l$ & 1.209 \\
        $A_{C}^{(2)}$  & $0.890$ \\
        $A_{C}^{(3)}$  & $-0.984$ \\
        \bottomrule
\end{tabular}
\caption{
    The different contributions of the four-pion and six-pion amplitudes (all taken in the flavor-stripped form analogous to Eq.~\eqref{eq:A6piP15}).
    LO indicates the lowest order and NLO next-to-leading order.
    The inputs are as given in the main text with $p=0.1\,\text{GeV}$.
    We quote only the real parts. Following Eq.~\eqref{eq:defA4}, $A^{(4)}(p_1,p_2,\dots,p_6)$ is the sum of terms in the bottom part of the table.
    The amplitudes are multiplied by a fitting power of $F_\pi$ consistently with the respective definitions (see, in particular, Eq.~\eqref{eq:defA4}) and (notably by $F_\pi^2$) to obtain dimensionless results.
}
    \label{tab:numerics}
\end{table}
Even though --- at least in this particular kinematical setting --- the individual contributions are relatively sizable, cancellations take place.
As a result, the overall contribution of the triangle functions is negligible compared to the $\bar J$ and polynomial parts, which clearly dominate in size also in terms of individual contributions.
This can be seen in Tab.~\ref{tab:numerics} comparing the left- and right-hand sides of the bottom part of the table.

In the limit $p\to0$, we find the following analytical expressions:
\begin{align}
&\hphantom{\operatorname{Re}\,\,}F_\pi^2A^{(2)}(p_1,p_2,\dots,p_6)\big|_{p\to0}
=5\,\frac{M_\pi^2}{F_\pi^2}\,,\\
\begin{split}
&F_\pi^2\operatorname{Re}A^{(4)}(p_1,p_2,\dots,p_6)\big|_{p\to0}\\
&=\frac{M_\pi^4}{F_\pi^4}\bigg\{
(-44+30N)\kappa
+(24)\kappa
-\frac{1}{18}(2+225N)\kappa\\
&-\frac16(14+75N)L
+(16l_1^\text{r}+56l_2^\text{r}+6l_3^\text{r}+20l_4^\text{r})\\
&+\frac12\kappa\bigl[-(30-9N)+(20)-(16)\bigr]
\bigg\}\,.
\end{split}
\label{eq:p0limit}
\end{align}
Numerically, $F_\pi^2A^{(2)}(p_1,p_2\dots,p_6)\big|_{p\to0}\approx11.33$ and $F_\pi^2\operatorname{Re}A^{(4)}(p_1,p_2\dots,p_6)\big|_{p\to0}\approx6.416$.
In the case of the NLO amplitude, the numerically dominant part stemming from $A_J^{(1)}$, $A_J^{(2)}$, $A_\pi$, $A_L$ and $A_l$ already yields 6.400, and in this exact order the listed contributions appear in the above expression, the last row of Eq.~\eqref{eq:p0limit} then being dedicated to the contributions of $A_C^{(1)}$, $A_C^{(2)}$ and $A_C^{(3)}$.

\section{Conclusions}
\label{sec:conclusions}

In this paper we calculated the pion mass, decay constant, the four-pion and six-pion amplitude to NLO in the massive O$(N)$ nonlinear sigma model.
In Sect.~\ref{sec:theory} we described the model and constructed the relevant NLO Lagrangian in analogy with the two(-quark)-flavor ChPT Lagrangian~\cite{Gasser:1983yg}.
The mass, decay constant and the four-pion amplitude agree for $N=3$ with Ref.~\cite{Gasser:1983yg} and with general-$N$ results of Refs.~\cite{Dobado:1994fd,Bijnens:2009zi,Bijnens:2010xg}.

Our main result is the six-pion amplitude.
We split it in one-particle reducible and irreducible parts; see Eq.~\eqref{eq:A6piparts}.
The reducible part is given in Eq.~\eqref{eq:A6pi4pi}, where we chose to employ the off-shell four-pion amplitude generalizing (beyond $N=3$) the amplitude given in Refs.~\cite{Bijnens:1995yn,Bijnens:1997vq}.
The irreducible part can be divided in a large number of subparts, each satisfying the expected permutation symmetries, as given in Eq.~\eqref{eq:defA4}, and the expressions are given explicitly in App.~\ref{app:6pi:res}.
The choice of triangle loop integrals with high symmetry allows for a fairly compact expression.

Some numerical results for one particular momentum configuration are presented in Sect.~\ref{sec:numerical}.
The NLO correction is sizable but not very large.

Work is in progress to combine our results with the methods for extracting three-body scattering from finite volume in lattice QCD. We expect that our result might also be of interest for the amplitude community.

\begin{acknowledgments}
This work is supported  in  part  by the Swedish Research Council grants contract numbers 2016-05996 and 2019-03779.
\end{acknowledgments}

\appendix
\section{Conventions for the loop integrals}
\label{app:integrals}

Throughout the paper we treat the momenta ($p_1,\,\dots,\,p_6$) as incoming.
We introduce the following combinations:
\begin{align}
    q_1&=p_1+p_2\,,&  q_2&=p_3+p_4\,,&  q_1&=p_5+p_6\,,\notag\\
    r_1&=p_1-p_2\,,&  r_2&=p_3-p_4\,,&  r_3&=p_5-p_6\,.
\end{align}

The functions we use to represent our results are the standard Passarino--Veltman one-loop integrals.
We specify some of them for completeness and to fix our notation.
The simpler integrals with one and two propagators read
\begin{align}
A_0&=\frac1i\int\frac{\text{d}^dr}{(2\pi)^d}
        \frac{1}{r^2-M^2}
  =M^2\kappa\frac1{\tilde\epsilon}-M^2 L\,,\notag\\
\begin{split}
B_0(q^2)&=\frac{1}{i}\int\frac{\text{d}^dr}{(2\pi)^d}
        \frac{1}{\left(r^2-M^2\right)\left[(r-q)^2-M^2\right]}\\
    &=\kappa\frac1{\tilde\epsilon}-\kappa-L+\bar J(q^2)\,,
\end{split}
\label{eq:AB}
\end{align}
with (as in Eq.~\eqref{eq:epstilde})
\begin{equation}
\frac1{\tilde\epsilon}
\equiv\frac{1}{\epsilon}-\gamma_\text{E}+\log4\pi-\log\mu^2+1\,.
\end{equation}
We also remind the reader that we set
\begin{align}
  \kappa&=\frac{1}{16\pi^2}\,,&
  L&=\kappa\log\frac{M^2}{\mu^2}\,.
\end{align}

In what follows we use for the Feynman denominators the compact notation
\begin{equation}
D(\pm q_i)
\equiv(r\mp q_i)^2-M^2\,,
\end{equation}
while setting $D_0\equiv D(0)=r^2-M^2$.
Then the integrals $A_0$ and $B_0$ from Eq.~\eqref{eq:AB} can be simply written as
\begin{align}
A_0&=\frac1i\int\frac{\text{d}^dr}{(2\pi)^d}
        \frac{1}{D_0}\,,\notag\\
\begin{split}
B_0(q_1^2)&=\frac{1}{i}\int\frac{\text{d}^dr}{(2\pi)^d}
        \frac{1}{D_0D(q_1)}\,.
\end{split}
\end{align}

It is the tensor triangle one-loop integrals of higher ranks which generate lengthy expressions upon reduction to the scalar ones.
Regarding the rank-3 integrals, the following combination has more symmetries than the first term only:
\begin{equation}
\begin{split}
&C_3(p_1,p_2,\dots,p_6)
=\frac{1}{3}\frac{1}{i}\int\frac{\text{d}^dr}{(2\pi)^d}
  \frac{r\cdot r_1 \,r\cdot r_2\, r\cdot r_3}{D_0}\\
&\times\Bigg[\frac{1}{D(q_1)D(-q_2)}+\frac{1}{D(q_2)D(-q_3)}+\frac{1}{D(q_3)D(-q_1)}\Bigg]\,.
\end{split}
\label{eq:C3}
\end{equation}
Moreover, note that, contrary to a naive (loop-momenta-)power counting applied to separate terms in Eq.~\eqref{eq:C3}, the combination $C_3$ is UV finite.
One should think of $C_3$ as a function of three pairs $(p_1,p_2)$, $(p_3,p_4)$, and $(p_5,p_6)$.
Such a combination differs from the first term alone only by terms with two or fewer $r\cdot r_i$.
One can see this in terms of a shift of the integration variable: One obtains the first propagator from the second one using $r\to r+q_2$, and from the last one using $r\to r-q_1$.
The combination $C_3(p_1,p_2,p_3,p_4,p_5,p_6)$ is antisymmetric under the following set of operations (taking each line separately):
\begin{gather}
  p_1\leftrightarrow p_2\,;\label{eq:P1}\\
 (p_1,p_2)\leftrightarrow(p_3,p_4)\,,\quad
 (p_1,p_2)\leftrightarrow(p_5,p_6)\label{eq:P2}\,.
\end{gather}
Hence, it is antisymmetric under the interchange of the momenta inside each pair and
antisymmetric under the interchange of two pairs.
Note that this generalization already follows from Eqs.~\eqref{eq:P1}, \eqref{eq:P2}.

One can also define symmetric combinations with fewer terms in the numerator (for rank-2 integrals):
\begin{align}
  C_{21}(p_1,p_2,\dots,p_6)
  &=\frac{1}{i}\int\frac{\text{d}^dr}{(2\pi)^d}
  \frac{r\cdot r_1 \,r\cdot r_2}
  {D_0D(q_1)D(-q_2)}\,,\notag\\
  C_{22}(p_1,p_2,\dots,p_6)
  &=\frac{1}{i}\int\frac{\text{d}^dr}{(2\pi)^d}\frac{r\cdot r_2 \,r\cdot r_3}
  {D_0D(q_2)D(-q_3)}\,,\notag\\
  C_{23}(p_1,p_2,\dots,p_6)
  &=\frac{1}{i}\int\frac{\text{d}^dr}{(2\pi)^d}
  \frac{r\cdot r_3 \,r\cdot r_1}
  {D_0D(q_3)D(-q_1)}\,.
\end{align}
Notice the cyclic symmetry.
To rewrite the results stemming directly from the diagrams in terms of the integrals $C_{2i}$, one needs to use the change of integration variables $r\to r+q_2$ and $r\to r-q_1$.
The last two integrals ($C_{22}$ and $C_{23}$) can be related to the first one ($C_{21}$) with pairs interchanged and pieces with at most one $r\cdot r_i$.
Thus we only need $C_{21}$ to express the final result.
Finally, $C_{21}$ is antisymmetric under the interchange $p_1\leftrightarrow p_2$ and symmetric under $(p_1,p_2)\leftrightarrow(p_3,p_4)$ and $p_5\leftrightarrow p_6$.

The integrals with one product $r\cdot r_i$ in the numerator can also be defined in a way symmetric under cyclic interchange:
\begin{align}
  C_{11}(p_1,p_2,\dots,p_6)
  &=\frac{1}{i}\int\frac{\text{d}^dr}{(2\pi)^d}
  \frac{r\cdot r_3}{D_0D(q_1)D(-q_2)}\,,\notag\\
  C_{12}(p_1,p_2,\dots,p_6)
  &=\frac{1}{i}\int\frac{\text{d}^dr}{(2\pi)^d}
  \frac{r\cdot r_1}{D_0D(q_2)D(-q_3)}\,,\notag\\
  C_{13}(p_1,p_2,\dots,p_6)
  &=\frac{1}{i}\int\frac{\text{d}^dr}{(2\pi)^d}
  \frac{r\cdot r_2}{D_0D(q_3)D(-q_1)}\,.
\end{align}
Using the change of integration variables $r\to r+q_2$ and $r\to r-q_1$, the last two integrals can be again related to the first one with pairs interchanged and parts without $r\cdot r_i$ in the numerator.
Thus we only need $C_{11}$ to express the final result.
The integral $C_{11}$ is antisymmetric under the interchange $p_5\leftrightarrow p_6$, $(p_1,p_2)\leftrightarrow(p_3,p_4)$, and is symmetric under $p_1\leftrightarrow p_2$ and $ p_3\leftrightarrow p_4$.

Finally, we define
\begin{equation}
  C(p_1,p_2,\dots,p_6)
  =\frac{1}{i}\int\frac{\text{d}^dr}{(2\pi)^d}\frac{1}{D_0D(q_1)D(-q_2)}\,,
\end{equation}
which is symmetric under $p_1\leftrightarrow p_2$ and under all pair interchanges.

We express the amplitude in terms of $C_3$, $C_{21}$, $C_{11}$ and $C$.
As already mentioned, the former three can be expressed in terms of $C$, but the expressions are cumbersome and lead to a very long expression for the amplitude.
We have therefore kept all these four, among which only $C_{21}$ has a UV infinite part:
\begin{equation}
\begin{split}
  &C_{21}(p_1,p_2,p_3,p_4,p_5,p_6)\\
  &=\kappa\,\frac{r_1\cdot r_2}{4}
  \frac1{\tilde\epsilon}+\overline C_{21}(p_1,p_2,p_3,p_4,p_5,p_6)\,.
\end{split}
\end{equation}  

A basis of momenta (up to the relation that is only valid in four dimensions)
which makes it easier to see symmetries is
\begin{align}
  q_1^2,\,q_2^2,\,q_3^2,\,r_1\cdot r_2,\,&\,r_2\cdot r_3,\,r_3\cdot r_1\,,\notag\\
  (q_1-q_2)\cdot r_3&=2q_1\cdot r_3\,,\notag\\
  (q_2-q_3)\cdot r_1&=2q_2\cdot r_1\,,\notag\\
  (q_3-q_1)\cdot r_2&=2q_3\cdot r_2\,.
\end{align}

\onecolumngrid

\section{The six-pion amplitude expressions}
\label{app:6pi:res}

In this section, when we refer to a pair we mean $(p_1,p_2)$, $(p_3,p_4)$ or $(p_5,p_6)$.
Note that in order to get the amplitude in this simpler form, we had to use symmetry properties of the integrals as well as many kinematic relations.
The notation $R_{ijklmn}$ indicates that in that term which it multiplies,
$\{p_1,p_2,p_3,p_4,p_5,p_6\}$ needs to be replaced by $\{p_i,p_j,p_k,p_l,p_m,p_n\}$.

Because of the symmetries, only one combination of $C_3$ with different arguments can emerge:
\begin{equation}
  A_{C_3}
  =\left(
  R_{152346}+R_{152436}+R_{162345}+R_{162435}
  -R_{132546}-R_{132645}-R_{142536}-R_{142635}
  \right) C_3(p_1,p_2,p_3,p_4,p_5,p_6)\,.
\end{equation}
For $\overline C_{21}$, two different cases appear.
One is when the last two of its
arguments (after the replacement operators $R$ are applied) correspond to one of the pairs:
\begin{equation}
  A_{C_{21}}^{(1)}
  =\left(R_{132456}+R_{142356}+R_{152634}+R_{162534}+R_{354612}+R_{364512}\right)
  \overline C_{21}(p_1,p_2,p_3,p_4,p_5,p_6)\left(4 p_5\cdot p_6+2M_\pi^2\right).
\end{equation}
The other combination is when none of the pairs shows up:
\begin{equation}
\begin{split}
A_{C_{21}}^{(2)}
&=\Big(
    R_{132546}+R_{132645}+R_{142536}+R_{142635}+R_{231546}+R_{241536}+R_{231645}+R_{241635}\\
  &-R_{234516}-R_{234615}-R_{243516}-R_{243615}+R_{253614}+R_{254613}+R_{263514}+R_{264513}\\
  &-R_{143526}+R_{163524}-R_{143625}+R_{153624}-R_{134526}+R_{164523}-R_{134625}+R_{154623}
\Big)\\
&\times\overline C_{21}(p_1,p_2,p_3,p_4,p_5,p_6)
    \left(\frac13(p_1+p_2)\cdot(p_5-p_6)-p_5\cdot  p_6\right).
\end{split}
\end{equation}
The terms with $C_{11}$ can be written as one combination
\begin{equation}
\begin{split}
A_{C_{11}}
&=\Big(
    R_{132546}+R_{132645}+R_{142536}+R_{142635}-R_{152346}-R_{152436}-R_{162345}-R_{162435}\\
  &+R_{234516}+R_{234615}+R_{243516}+R_{243615}+R_{253614}+R_{254613}+R_{263514}+R_{264513}\\
  &-R_{351426}-R_{351624}-R_{361425}-R_{361524}-R_{451326}-R_{451623}-R_{461325}-R_{461523}
  \Big)\\
&\times C_{11}(p_1,p_2,p_3,p_4,p_5,p_6)
    \big[(p_1\cdot  p_2)(p_3\cdot p_4)\big]\,.
\end{split}
\end{equation}
The terms containing $C$ are of three types: with three, one or none of the pairs.
The part with three pairs is
\begin{equation}
\begin{split}
A_C^{(1)}
 &=C(p_1,p_2,p_3,p_4,p_5,p_6)
 \bigl[-N M_\pi^6-2(N-1)M_\pi^4(p_1\cdot p_2+p_3\cdot p_4+p_5\cdot p_6)\bigr.\\
  &-\bigl.4(N-2)M_\pi^2(p_1\cdot p_2\,p_3\cdot p_4+p_3\cdot p_4\,p_5\cdot p_6+p_5\cdot p_6\,p_1\cdot p_2)
  -8(N-3)p_1\cdot p_2\,p_3\cdot p_4\,p_5\cdot p_6\bigr].
\end{split}
\end{equation}
Those with one pair are
\begin{equation}
\begin{split}
A_C^{(2)}
&=\Big(R_{123546}+R_{123645}+R_{341526}+R_{341625}+R_{561324}+R_{561423}\Big)\\
&\times C(p_1,p_2,p_3,p_4,p_5,p_6)
    \left(-4 p_1\cdot p_2-2M_\pi^2\right)p_3\cdot p_4\,p_5\cdot p_6\,.
\end{split}
\end{equation}
The case when no pair is present in the arguments of $C$s reads
\begin{equation}
\begin{split}
A_C^{(3)}
 &=\Big(R_{132546}+R_{132645}+R_{142536}+R_{142635}+R_{152364}+R_{152463}+R_{162354}+R_{162453}\Big)\\
 &\times C(p_1,p_2,p_3,p_4,p_5,p_6)
  \Big[p_1\cdot p_2\,p_3\cdot p_4\,p_5\cdot p_6
  -p_1\cdot p_2\,p_3\cdot p_4\,(p_1+p_2)\cdot(p_5-p_6)\\
 &-p_3\cdot p_4\,p_5\cdot p_6\,(p_3+p_4)\cdot(p_1-p_2)
  +p_5\cdot p_6\,p_1\cdot p_2\,(p_5+p_6)\cdot(p_3-p_4)\Big]\,.
\end{split}
\end{equation}
The terms containing $\bar J$ are similarly split in two expressions depending on whether the argument of $\bar J$ corresponds to a pair or not:
\begin{align}
\begin{split}
 A_J^{(1)}
 &=\Bigl(1+R_{341256}+R_{561234}\Bigr)\bar J\left((p_1+p_2)^2\right)
  \Big[(2N-3)(p_1\cdot p_2)^2+(2N-7)p_1\cdot p_2(p_3\cdot p_4+p_5\cdot p_6)\\
  &+\left(4N-\frac{13}2\right)M_\pi^2p_1\cdot p_2+\left(N-\frac52\right)M_\pi^2(p_3\cdot p_4+p_5\cdot p_6)
  +\frac32(N-1)M_\pi^4\Big]\,,
\end{split}\\
\begin{split}
 A_J^{(2)}
  &=\Bigl(
       R_{132456}+R_{142356}+R_{231456}+R_{241356}+R_{152634}+R_{251634}\\
       &+R_{162543}+R_{261543}+R_{536412}+R_{635421}+R_{546312}+R_{645321}
     \Bigr)\\
  &\times\bar J\left((p_1+p_2)^2\right)
    \bigg\{\frac34\,p_5\cdot p_6\,(p_5\cdot p_6-p_1\cdot p_3-p_2\cdot p_4)
    -\frac38 M_\pi^2 (p_1\cdot p_3+p_2\cdot p_4+3 p_5\cdot p_6)\\
  &+\frac1{24}(p_1-p_3)\cdot(p_2-p_4)\big[2 (p_1\cdot p_2-p_3\cdot p_4)
   +19 p_5\cdot p_6-p_1\cdot p_3- p_2\cdot p_4+4 M_\pi^2\big]\\
  &+\frac1{24}\left[(p_1-p_3)\cdot(p_2-p_4)\right]^2
   +\frac12(p_1\cdot p_2-p_3\cdot p_4) (3 p_5\cdot p_6+M_\pi^2)
   -\frac58 M_\pi^4\bigg\}\,.
\end{split}
\end{align}
The polynomial part is finally
\begin{align}
\begin{split}
A_\pi
&=\kappa\bigg[
    \bigg(\frac{49}{144}-\frac N2\bigg) (q_1^4+q_2^4+q_3^4)
    +\bigg(\frac{281}{72}-N\bigg) (q_1^2 q_2^2+q_2^2 q_3^2+q_3^2 q_1^2)
    -\bigg(\frac{217}{18}-3N\bigg) M_\pi^2 (q_1^2+q_2^2+q_3^2)\\
    &-\frac{5}{36}\big[(r_1\cdot r_2)^2+(r_2\cdot r_3)^2+(r_3\cdot r_1)^2\big]
    +\bigg(23-\frac92N\bigg) M_\pi^4
\bigg]\,,
\end{split}\\
\begin{split}
A_L
&=L\bigg[
    \bigg(\frac13-\frac N2\bigg) (q_1^4+q_2^4+q_3^4)
    +\bigg(\frac{13}3-N\bigg) (q_1^2 q_2^2+q_2^2 q_3^2+q_3^2 q_1^2)
    -\bigg(\frac{41}3-3N\bigg) M_\pi^2 (q_1^2+q_2^2+q_3^2)\\
    &-\frac16\big[(r_1\cdot r_2)^2+(r_2\cdot r_3)^2+(r_3\cdot r_1)^2\big]
    +\bigg(27-\frac92 N\bigg) M_\pi^4\bigg]\,,
\end{split}\\
\begin{split}
A_l
&=\frac12(4 l_1^\text{r}+l_2^\text{r}) (q_1^2+q_2^2+q_3^2)^2
    -(20l_1^\text{r}+l_2^\text{r}-4 l_4^\text{r}) M_\pi^2 (q_1^2+q_2^2+q_3^2)\\
    &+2l_2^\text{r} \big[(q_1\cdot r_3)^2+(q_2\cdot r_1)^2+(q_3\cdot r_2)^2\big]
    +6(8 l_1^\text{r}+l_3^\text{r}-2 l_4^\text{r}) M_\pi^4\,.
\end{split}
\end{align}
 
\twocolumngrid

\renewcommand{\raggedright}{}
%\bibliography{references}

\providecommand{\href}[2]{#2}\begingroup\raggedright\endgroup

\end{document}